\documentclass[a4paper,fleqn,usenatbib]{mnras}

\usepackage[T1]{fontenc}
\usepackage{ae,aecompl}

\usepackage{graphicx}	
\usepackage{amsmath}	
\usepackage{amssymb}	

\usepackage{hyperref}
\usepackage{xspace}
\usepackage{rotating}
\usepackage{ulem}
\usepackage[flushleft]{threeparttable}

\newcommand{\masyr}{mas yr$^{-1}$\xspace}

\title[Arches and Quintuplet absolute PMs]{The absolute proper motions of the Arches and Quintuplet clusters}

\author[M. Libralato et al.]{
  Mattia Libralato$^{1}$\thanks{E-mail: \href{mailto:libra@stsci.edu}{libra@stsci.edu}},
  Mark Fardal$^{1}$,
  Daniel Lennon$^{2,3}$,
  Roeland P. van der Marel$^{1,4}$ and
  \newauthor
  Andrea Bellini$^{1}$ \\~\\
$^{1}$Space Telescope Science Institute 3700 San Martin Drive, Baltimore, MD 21218, USA \\
$^{2}$Instituto de Astrof\'isica de Canarias, E-38205 La Laguna, Tenerife, Spain \\
$^{3}$Universidad de La Laguna, Dpto. Astrof\'isica, E-38206 La Laguna, Tenerife, Spain \\
$^{4}$Center for Astrophysical Sciences, Department of Physics \& Astronomy, Johns Hopkins University, Baltimore, MD 21218, USA
}

\date{Accepted 2020 August 03. Received 2020 August 03; in original form 2020 January 17}

\pubyear{2020}

\begin{document}
\label{firstpage}
\pagerange{\pageref{firstpage}--\pageref{lastpage}}
\maketitle

\begin{abstract}
  Arches and Quintuplet are two young, massive clusters projected near
  the Galactic Center. To date, studies focused on understanding their
  origin have been based on proper motions (PMs) derived in the
  clusters' reference frames and required some assumptions about their
  3D motion. In this paper, we combine public PM catalogs of these
  clusters with the Gaia DR2 catalog and, for the first time,
  transform the relative PMs of the Arches and Quintuplet clusters
  onto an absolute reference system. We find that the absolute PM of
  the Arches is $(\mu_\alpha \cos\delta,\mu_\delta)$ $=$
  $(-1.45\ \pm\ 0.23,-2.68\ \pm\ 0.14)$ \masyr, and that of the
  Quintuplet is $(\mu_\alpha \cos\delta,\mu_\delta)$ $=$
  $(-1.19\ \pm\ 0.09,-2.66\ \pm \ 0.18)$ \masyr. These values suggest
  that these systems are moving almost parallel to the Galactic
  plane. A measurement of the clusters' distances is still required to
  meaningfully constrain the clusters' orbits and shed light on the
  origin of the Arches and Quintuplet.
\end{abstract}



\defcitealias{2015A&A...578A...4S}{S15}


\section{Introduction}

The Galactic Center (GC) represents a unique ecosystem in our
Galaxy. Indeed, despite the harsh environment around the supermassive
black hole Sgr A*, at least three young, massive clusters and several
massive isolated stars are present in the region. The recent star
formation suggested by the presence of these objects is still a
conundrum. Either ``in-situ'' or ``accreted'' formation scenarios have
been proposed so far, but the observational pieces of information at
our disposal cannot firmly rule out either of these theories
\citep[see, e.g., the review of][and references
  therein]{2010RvMP...82.3121G}.

Two of the most studied among these young objects are the Arches and
Quintuplet clusters. Arches and Quintuplet \citep[$\sim$2.5 Myr and
  $\sim$4 Myr,
  respectively][]{2004ApJ...611L.105N,1999ApJ...514..202F} are massive
\citep[$\gtrsim10^4$;
  e.g.,][]{1999ApJ...525..750F,2010MNRAS.409..628H} clusters located
close to the Galactic plane at a projected distance of about 20--30 pc
from the GC. Where these clusters formed and how they have survived in
such a dense region is, however, still a puzzle.

In order to find the exact birth sites of the Arches and Quintuplet
clusters, it is necessary to compute their orbits backward in
time. The computation of an orbit requires six kinematic coordinates:
position, distance, proper motion (PM) and radial velocity (RV). For
these clusters, three of them (position and RV) are known. While the
distance is still the most difficult observable piece of information
to obtain, the PMs of Arches and Quintuplet clusters have been
measured in the past by different authors \citep[e.g.,][and references
  therein]{2012ApJ...751..132C,2014ApJ...789..115S,2015A&A...578A...4S,2019ApJ...870...44H,2019ApJ...877...37R}.

The orbits of the Arches and Quintuplet clusters presented to date
are, however, based on the relative PM between cluster stars and field
objects, and are based on the assumption that field stars are on
average at rest with respect to the GC. Furthermore, the field-star
distribution in the vector-point diagram (VPD) is often modeled with a
single 2D-Gaussian distribution, a representation that cannot properly
describe the true, complex kinematic scene in the GC region. These
assumptions can result in large systematic errors and lack of
accuracy.

In this paper, we take advantage of the Gaia Data Release 2
\citep[DR2,][]{2016A&A...595A...1G,2018A&A...616A...1G} catalog and
compute the first estimate of the absolute PMs of the Arches and
Quintuplet clusters by transforming the public relative PMs of
\citet[][hereafter S15]{2015A&A...578A...4S} onto an absolute
reference frame. We also briefly investigate some orbits implied by
the absolute PMs of these objects, and discuss their possible origin
site (Appendix \ref{orbit}).

\section{Data sets}\label{data}

We made use of the catalogs of \citetalias{2015A&A...578A...4S}, to
which we refer for a complete description of the data reduction. PMs
were derived by means of multi-epoch (3--5 yr of temporal baseline)
$K_{\rm S}$-filter images obtained with the NAOS-CONICA (NACO) system
at the VLT. PMs were computed relative to the bulk motion of the
clusters, i.e., the average motion of cluster stars in the VPD is
consistent with zero, while Bulge and Disk stars are located in
different parts of the VPD\footnote{We compared the VPDs obtained with
  the catalogs of \citetalias{2015A&A...578A...4S} with VPDs presented
  in other papers focused on the Arches and Quintuplet clusters
  \citep[e.g.,][]{2019ApJ...870...44H,2019ApJ...877...37R} and with
  those obtained from the Gaia DR2 catalog. We found that the PM
  distribution of the field stars along the $\alpha \cos\delta$
  direction obtained with the \citetalias{2015A&A...578A...4S}
  catalogs has an opposite sign from what is expected for Disk/Bulge
  stars in these fields. Therefore, we changed the sign of $\mu_\alpha
  \cos\delta$ in the catalogs of \citetalias{2015A&A...578A...4S}.}.

As described in \citetalias{2015A&A...578A...4S}, all images of a
given epoch were initially combined with the drizzle method
\citep{2002PASP..114..144F} without applying any geometric distortion
correction. The positions of the stars measured in these stacked
images of each epoch were then transformed onto the reference frame of
the first epoch by means of second-order polynomial functions, which
should solve for the relative rotation, offset, scale and distortion
between the frames.

\citet{2014PhDT.......214H} and \citetalias{2015A&A...578A...4S}
stated that there is not a uniform distortion solution for the NACO
detector because it depends on the isoplanatic angle and on the
adaptive-optics correction. According to
\citetalias{2015A&A...578A...4S}, the relative astrometric
uncertainties were minimized in their paper thanks to the small-dither
offsets between images and by keeping the same observational setup in
all epochs. Furthermore, the usage of second-order polynomials to
transform the positions between frames had probably absorbed part of
the distortion.\looseness=-4

\citet{2015MNRAS.453.3234P} solved for the distortion of NACO detector
in the same configuration as that used by
\citetalias{2015A&A...578A...4S}. The typical distortion of this
camera is of 0.2 pixel ($\sim 5$ mas), but it can be as large as 0.7
pixel ($\sim 19$ mas) in the lower-left corner of the detector.

The analysis and correction of possible systematics in the PMs of
\citetalias{2015A&A...578A...4S} is not straightforward. The
transformations between frames were computed by using cluster
stars. This means that we need to analyze the PM of cluster members to
detect possible systematic errors in the PMs. However, the
identification of cluster stars in the VPD is not always unambiguous,
especially in the outer fields of the two clusters where most of the
stars considered in the analysis in Sect.~\ref{gaia} are
located. Therefore, we chose to compute the absolute PMs of Arches and
Quintuplet clusters by using the original PMs of
\citetalias{2015A&A...578A...4S}.

Nevertheless, in Appendix~\ref{appsys} we examine the PMs of
\citetalias{2015A&A...578A...4S}, search for the presence of spatial-
and/or magnitude-dependent systematics, and discuss various approaches
to correct them. We find that the different corrections that can be
applied do not affect the value of the absolute PMs of the clusters.

\section{The absolute PMs of Arches and Quintuplet clusters}\label{gaia}

Relative PMs can be directly converted into an absolute reference
system by zero-pointing them to the PMs of very-distant objects like
quasars or galaxies \citep[e.g.,][]{2018ApJ...854...45L}, or
indirectly by relying on external catalogs. The former approach is not
feasible toward the GC because of the high extinction.

\begin{figure*}
  \centering
  \includegraphics[width=0.8\textwidth]{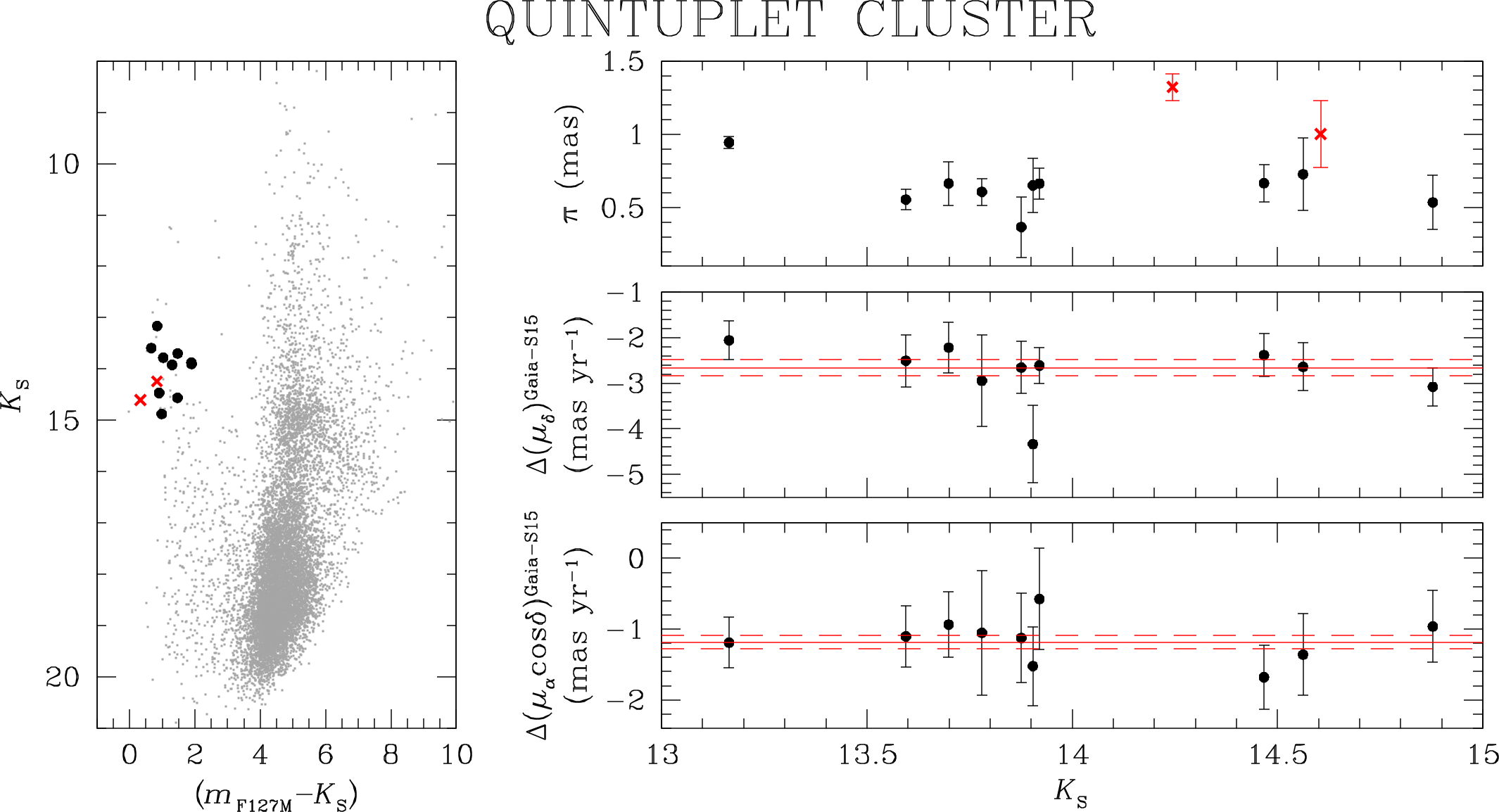}
  \caption{The $K_{\rm S}$ as a function of ($m_{\rm F127M}-K_{\rm
      S}$) CMD (left panel), in which we show all stars in the PM
    catalog of \citetalias{2015A&A...578A...4S} as gray points. Black
    dots and red crosses represent the few well-measured (see the text
    for details) Disk stars in common between the
    \citetalias{2015A&A...578A...4S} and Gaia-DR2 catalogs with a
    parallax smaller or larger than 1 mas, respectively. In the
    top-right panel, we plot the parallax $\pi$ (with error bars) from
    the Gaia DR2 catalog as a function of the $K_{\rm S}$ magnitude
    from the \citetalias{2015A&A...578A...4S} catalog. Only stars with
    a parallax smaller than 1 mas (distance from the Sun greater than
    1 kpc) are used to compute the absolute PM of the cluster. In the
    bottom and middle panels, we present the PM difference along
    $\alpha\cos\delta$ and $\delta$ directions for the Disk stars in
    common between the \citetalias{2015A&A...578A...4S} and Gaia-DR2
    catalogs. The error bars are the sum in quadrature of the Gaia and
    \citetalias{2015A&A...578A...4S} PM errors. The average PM
    differences are shown as red, solid lines; the corresponding
    errors as red, dashed lines.}
  \label{fig:abspmq}
  \vspace{0.5cm}
  \includegraphics[width=0.8\textwidth]{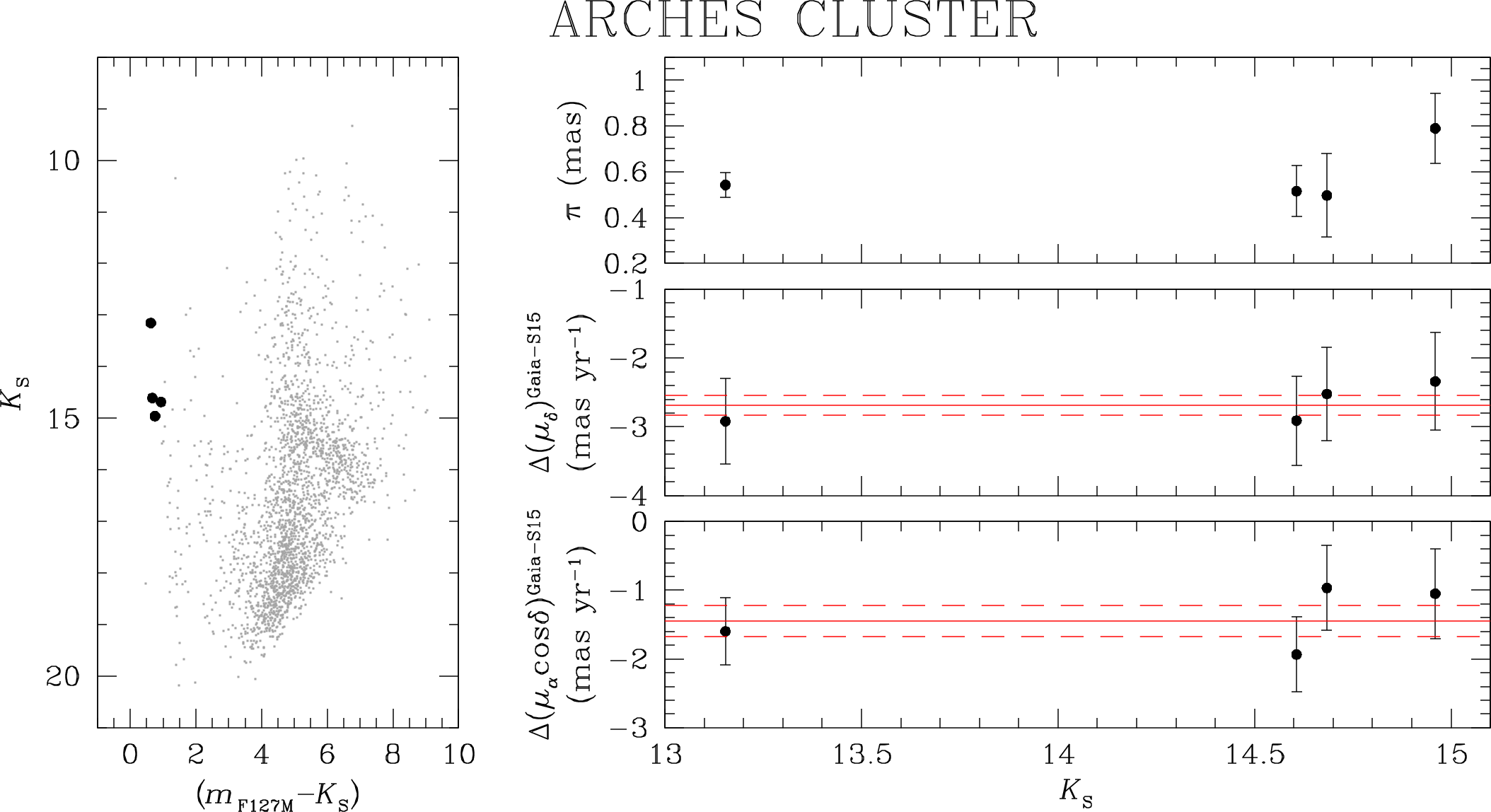}
  \caption{As in Fig.~\ref{fig:abspmq}, but for the Arches PM
    catalog.}
  \label{fig:abspma}
\end{figure*}

We derived the first estimate of the absolute PMs of the Arches and
Quintuplet clusters by linking the relative PMs of
\citetalias{2015A&A...578A...4S} to an absolute reference system using
the Gaia DR2 PMs. The depth of the Gaia catalog toward regions of high
extinction like the GC is a few kpc, which means that the only stars
in common between \citetalias{2015A&A...578A...4S} and Gaia are Disk
stars (as shown in the CMDs in Figs.~\ref{fig:abspmq} and
\ref{fig:abspma}).

We considered in the analysis stars in the Gaia DR2 catalog that have
a PM error in both coordinates lower than 1.0 \masyr and are brighter
than $G = 18.5$. We excluded stars in the
\citetalias{2015A&A...578A...4S} catalogs that are brighter than
$K_{\rm S} = 13$ (very bright stars suffer from non-linearity effects
in the NACO data), fainter than $K_{\rm S} = 17.5$ (the same magnitude
threshold adopted by \citetalias{2015A&A...578A...4S} to select
reference stars in the PM computation) or have a PM error greater than
1.0 mas yr$^{-1}$ in either coordinates. For the Arches cluster, we
also excluded all stars in their outermost field ``2'', which is not
part of the NACO mosaic, where the number of cluster stars is low and
the high fraction of field stars might have biased the computation of
the relative PMs. After these selections, we ended up with four and
twelve Gaia stars in common with the Arches and Quintuplet catalogs,
respectively.

Finally, to keep the two samples as homogeneous as possible, we
additionally excluded stars with a parallax larger than 1 mas
(distance from the Sun smaller than 1 kpc). The surviving four and ten
stars in the Arches and Quintuplet catalogs, respectively, have a
distance from the Sun between 1 and 2.8 kpc (see top-right panels in
Figs.~\ref{fig:abspmq} and \ref{fig:abspma}).

We computed the 3.5$\sigma$-clipped weighted average difference
between Gaia and \citetalias{2015A&A...578A...4S} PMs (middle- and
bottom-right panels in Figs.~\ref{fig:abspmq} and
\ref{fig:abspma}). The weights are defined as the sum in quadrature of
Gaia and \citetalias{2015A&A...578A...4S} PM errors. Since the PMs of
\citetalias{2015A&A...578A...4S} are relative to the bulk motion of
the clusters, these Gaia-\citetalias{2015A&A...578A...4S}
weighted-average PM differences are the absolute PMs of the Arches and
Quintuplet clusters. We find:
\begin{equation}
  \begin{gathered}
    (\mu_\alpha \cos\delta,\mu_\delta)^{\rm Arches} \\
    = \\
    (-1.45\ \pm\ 0.23,-2.68\ \pm\ 0.14) \textrm{ \masyr,}
  \end{gathered}
  \label{eq:1}
\end{equation}
\begin{equation}
  \begin{gathered}
    (\mu_\alpha \cos\delta,\mu_\delta)^{\rm Quintuplet} \\
    = \\
    (-1.19\ \pm\ 0.09,-2.66\ \pm \ 0.18) \textrm{ \masyr.}
  \end{gathered}
  \label{eq:2}
\end{equation}
These PMs represent the first estimate of the absolute PMs of the
Arches and Quintuplet clusters. The quoted errors are the standard
errors of the mean of the Gaia-\citetalias{2015A&A...578A...4S} PM
difference. We have not included the effects of the systematic errors
in the Gaia DR2, as these are significantly smaller than the
statistical errors calculated here \citep{2018A&A...616A...2L}.

\begin{figure*}
  \centering
  \includegraphics[width=0.9\textwidth]{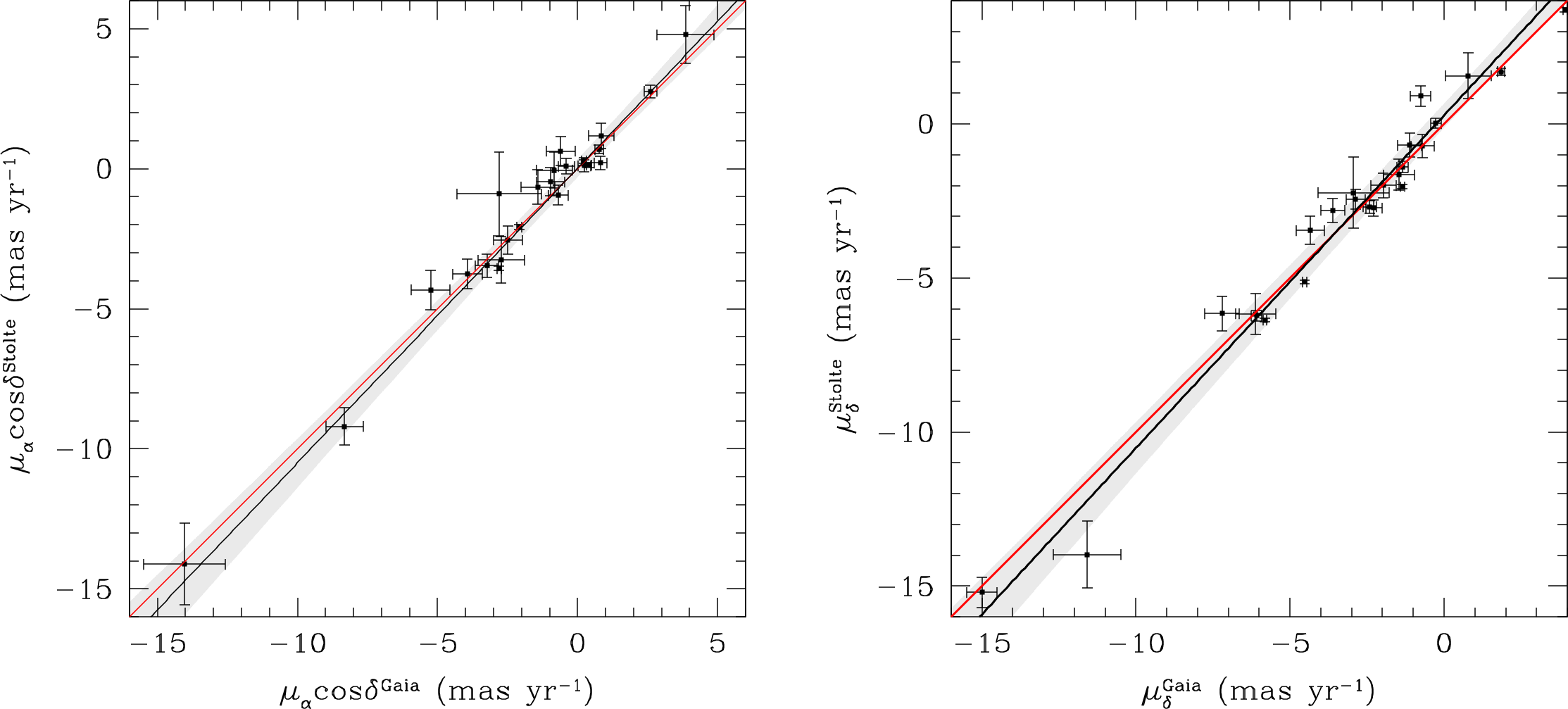}
  \caption{Comparison between the absolute PMs of
    \citetalias{2015A&A...578A...4S} and the Gaia-DR2 PMs along
    $\alpha\cos\delta$ (left panel) and $\delta$ (right panel)
    directions for the Quintuplet cluster. All stars in common between
    the two catalogs are shown. In each panel, the red line is the
    plane bisector. The black line represents the best fit to the
    point obtained taking into account for the errors in both
    coordinates. The gray area is the 1$\sigma$ confidence region.}
  \label{fig:gaiaq}
  \vspace{0.5cm}
  \includegraphics[width=0.9\textwidth]{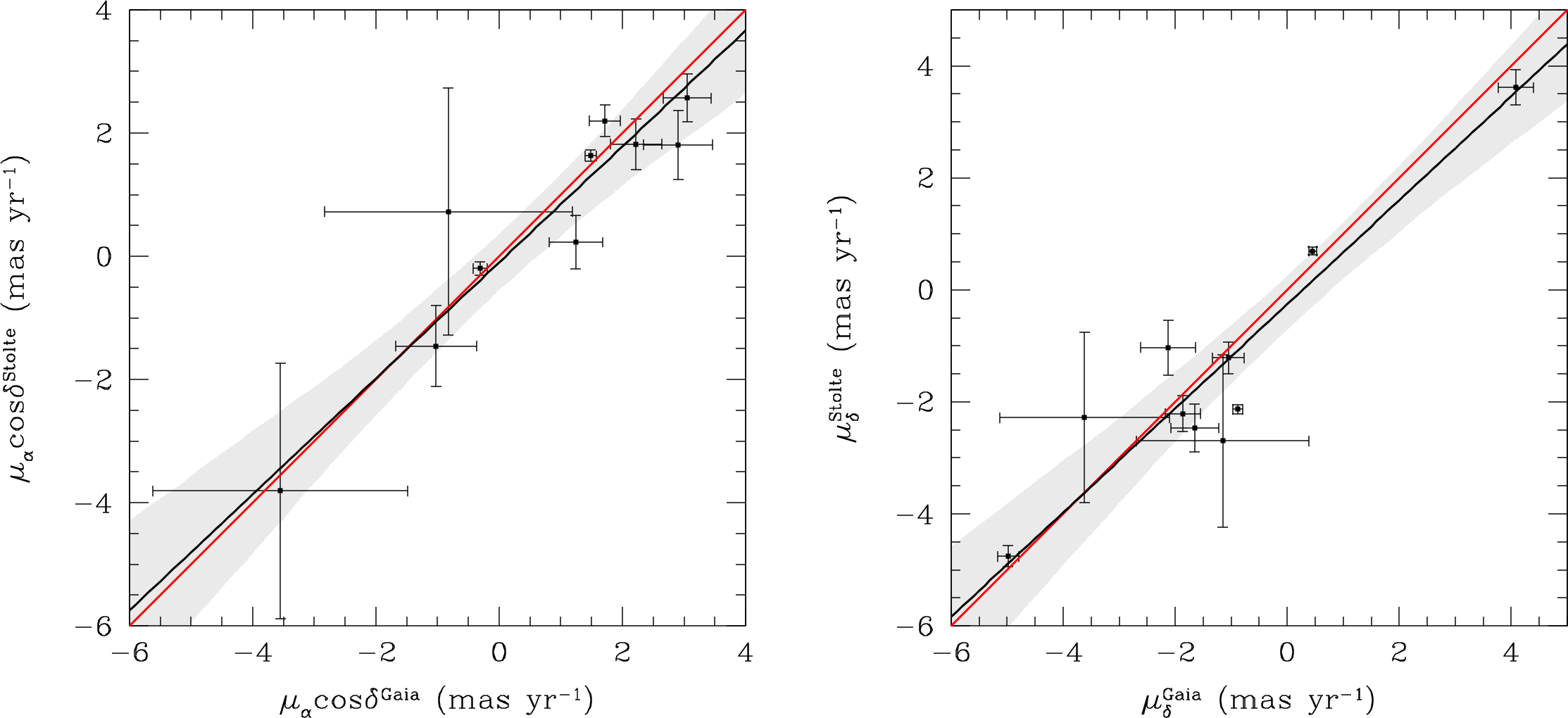}
  \caption{As in Fig.~\ref{fig:gaiaq}, but for the Arches PM
    catalog. All stars in common between the two catalogs are shown,
    except for those in the outermost field ``2'' of the Arches
    mosaic.}
  \label{fig:gaiaa}
\end{figure*}

In Fig.~\ref{fig:gaiaq} and \ref{fig:gaiaa} we show a comparison
between the PMs of the Gaia DR2 catalog and those of
\citetalias{2015A&A...578A...4S} after they have been converted to
absolute values. All stars in common between these catalogs were used
in the comparison (except for those in the outermost field ``2'' of
the Arches mosaic). The red line represents the plane bisector. The
black line is the best fit to the points, the gray region is the
1$\sigma$-error region. The two sets of PMs are in agreement at the
1$\sigma$ level, thus further confirming the reliability of the PMs of
\citetalias{2015A&A...578A...4S} even without a geometric-distortion
correction of the NACO data.

We converted the absolute PMs of the Arches and Quintuplet clusters
from Equatorial to Galactic coordinates. The Equatorial-to-Galactic
conversion of the PMs is represented by a rotation matrix, but the
exact transformation of the PM-error ellipses is not as
trivial. Therefore, we followed a Monte Carlo approach. For each
cluster, we used 10\,000 samples and assumed Gaussian distributions
for $\mu_\alpha \cos\delta$ and $\mu_\delta$ with average and $\sigma$
equal to the absolute PMs and errors in Equatorial coordinates. We
then converted the PMs of these samples to the Galactic reference
system and defined as the best estimate and uncertainties the average
value and the standard deviation of the obtained distributions,
respectively. The resulting absolute PMs in Galactic coordinates are:
\begin{equation}
  \begin{gathered}
    (\mu_{\rm l}\cos{\rm b},\mu_{\rm b})^{\rm Arches} \\
    = \\
    (-3.05\ \pm\ 0.17,-0.16\ \pm\ 0.20) \textrm{ \masyr,}
  \end{gathered}
  \label{eq:3}
\end{equation}
\begin{equation}
  \begin{gathered}
    (\mu_{\rm l}\cos{\rm b},\mu_{\rm b})^{\rm Quintuplet} \\
    = \\
    (-2.89\ \pm\ 0.16,-0.38\ \pm \ 0.12) \textrm{ \masyr.}
  \end{gathered}
  \label{eq:4}
\end{equation}

\subsection{Comparison with the literature}

Even though there are no previous estimates of the absolute PMs for
the Arches and Quintuplet clusters, we can still make a qualitative
comparison with the literature by assuming that the Bulge stellar
components identified in other works have mean zero velocity with
respect to the GC.

The PM of Sgr A* is
$(\mu_\alpha \cos\delta,\mu_\delta)^{\rm Sgr A*} = (-3.156,-5.585)$
mas yr$^{-1}$ \citep{2020ApJ...892...39R}. From (\ref{eq:1}) and
(\ref{eq:2}), we have that the PMs of the Arches and Quintuplet
clusters with respect to Sgr A* are:\looseness=-4
\begin{equation}
  \begin{gathered}
    (\mu_\alpha \cos\delta,\mu_\delta)^{\rm Arches-Sgr A^*} \\
    = \\
    (1.71\ \pm\ 0.23,2.91\ \pm\ 0.14) \textrm{ \masyr,}
  \end{gathered}
  \label{eq:5}
\end{equation}
\begin{equation}
  \begin{gathered}
    (\mu_\alpha \cos\delta,\mu_\delta)^{\rm Quintuplet-Sgr A^*} \\
    = \\
    (1.97\ \pm\ 0.09,2.93\ \pm\ 0.18) \textrm{ \masyr.}
  \end{gathered}
  \label{eq:6}
\end{equation}
We did not include the PM errors of Sgr A* in the error budget because
they are too small to be significant.

Several estimates of the Arches/Quintuplet PMs have been made over the
years. Here we compare the most recent estimates of the PMs for each
cluster, which are based on completely different data sets from those
of \citetalias{2015A&A...578A...4S} and previous papers from the same
authors \citep{2008ApJ...675.1278S,2014ApJ...789..115S}.

\citet{2019ApJ...870...44H} and \citet{2019ApJ...877...37R} recently
computed the relative PMs of the Arches and Quintuplet clusters,
respectively, using \textit{Hubble Space Telescope} (\textit{HST})
data. The authors modeled the distribution of the field stars in the
VPD as the sum of different Gaussian distributions, thus obtaining a
sophisticated and reliable representation of the motion of the Bulge
stars in these fields. If we assume that the Bulge stars are, on
average, at rest with respect to Sgr A*, this means that the relative
PM of the clusters with respect to Sgr A* is given by the average PM
of Bulge stars in the relative-PM VPD, with opposite sign.

Looking at Fig.~4 of \citet{2019ApJ...870...44H}, the cyan Gaussian
(the field Gaussian 3) seems to fairly represent the PM distribution
of Bulge stars in the VPD of the Arches cluster. Hence, the relative
PM of the Arches cluster with respect to Sgr A* is defined by the
center of Gaussian 3 in their Table~7, with opposite sign:
\begin{equation}
  \begin{gathered}
    (\mu_\alpha \cos\delta,\mu_\delta)^{\rm Arches-Sgr A^*\ H19} \\
    = \\
    (1.90\ \pm\ 0.08,2.89\ \pm\ 0.10) \textrm{ \masyr.}
  \end{gathered}
  \label{eq:7}
\end{equation}
This value is in agreement with our independent estimate given in
(\ref{eq:5}).

We made a similar computation for the Quintuplet cluster using the
details provided in \citet{2019ApJ...877...37R}. The center of the
field stars in the VPD shown in their Fig.~7 seems to be between the
centers of the blue and cyan Gaussians (the field Gaussians 1 and
3). By means of the values given in their Table~2, the resulting
relative PM of the Quintuplet cluster with respect to Sgr A* is:
\begin{equation}
  \begin{gathered}
    (\mu_\alpha \cos\delta,\mu_\delta)^{\rm Quintuplet-Sgr A^*\ R19} \\
    = \\
    (1.99\ \pm\ 0.14,3.12\ \pm\ 0.18) \textrm{ \masyr.}
  \end{gathered}
  \label{eq:8}
\end{equation}
This estimate is also consistent with our estimate in (\ref{eq:6}) at
the $1\sigma$ level.

\begin{figure*}
  \centering
  \includegraphics[width=0.9\textwidth]{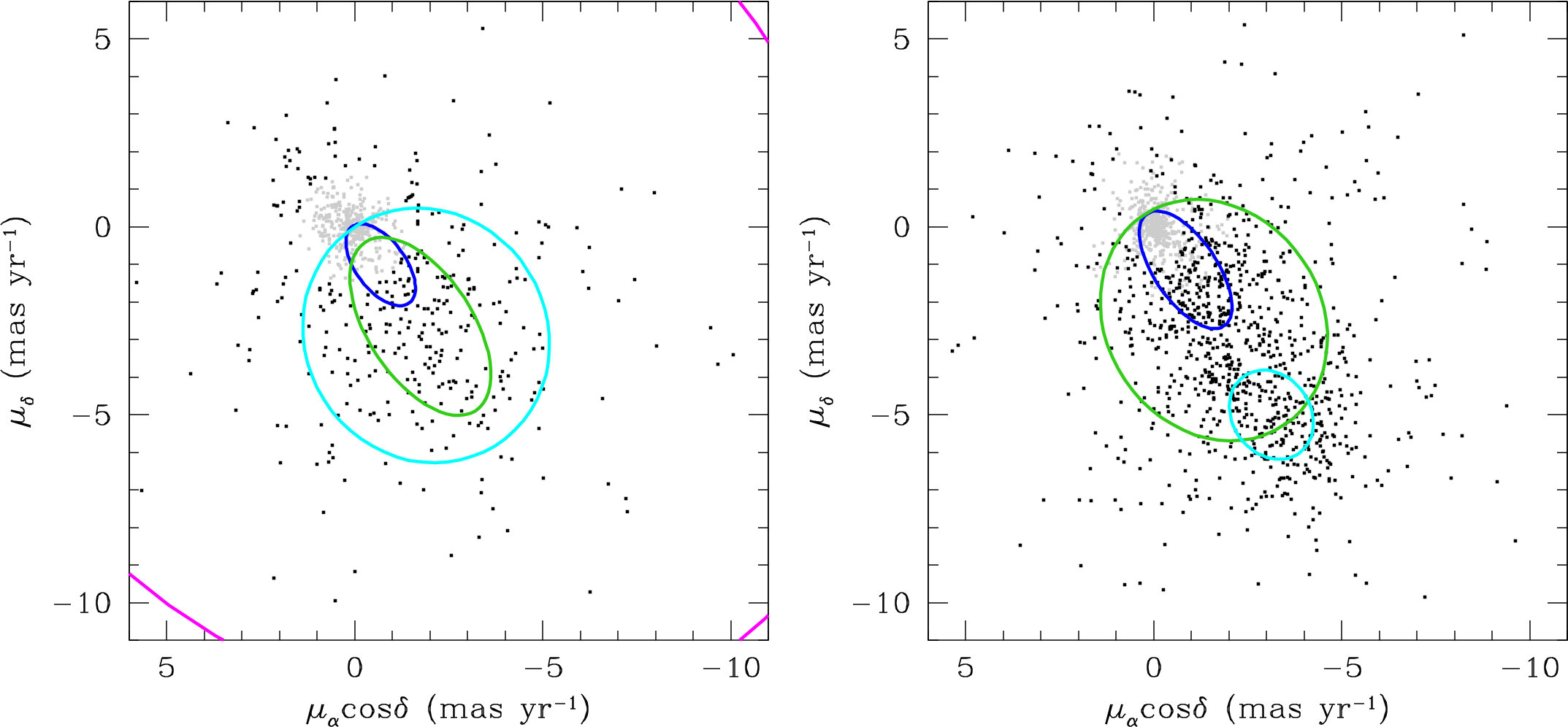}
  \caption{The VPD of the relative PMs of
    \citetalias{2015A&A...578A...4S} for the Arches (left panel) and
    Quintuplet (right panel) clusters. Gray points are bright
    ($K_{\rm S}<16$, i.e., stars with small PM errors) cluster stars
    defined according to the membership flag in
    \citetalias{2015A&A...578A...4S} catalogs. Bright ($K_{\rm S}<16$)
    stars redder than $(m_{\rm F127M}-K_{\rm S}) \sim 4$ are
    considered Bulge objects and are shown as black points. The
    $1\sigma$ Gaussian contours defined in \citet{2019ApJ...870...44H}
    for the Arches cluster (left panel) and in
    \citet{2019ApJ...877...37R} for the Quintuplet cluster (right
    panel) are shown as ellipses, color-coded as in the corresponding
    papers.}
  \label{fig:compqa}
\end{figure*}

For the sake of completeness, Fig.~\ref{fig:compqa} shows the VPDs of
the relative PMs of \citetalias{2015A&A...578A...4S} for the Arches
(left panel) and Quintuplet (right panel) catalogs. Black points
represent a sample of bright Bulge stars selected according to their
location in the $K_{\rm S}$ versus $(m_{\rm F127M}-K_{\rm S})$
color-magnitude diagram, i.e., redder than $(m_{\rm F127M}-K_{\rm S})
\sim 4$ (see, e.g., the CMDs in Figs.~\ref{fig:abspmq} and
\ref{fig:abspma}). The ellipses depicted in these plots represent the
$1\sigma$ Gaussian contours defined in \citet{2019ApJ...870...44H} for
the Arches cluster and in \citet{2019ApJ...877...37R} for the
Quintuplet cluster.

\section{Conclusions}

We took advantage of the public PM catalogs of
\citetalias{2015A&A...578A...4S} and of the Gaia DR2 to compute the
first estimate of the absolute PMs of the Arches and Quintuplet
clusters. One of the main advantages of PMs is that they enable orbit
calculations. We explore this topic in Appendix~\ref{orbit} through
approximate calculations in an axisymmetric potential. This provides
some insights into plausible dynamical histories and origins of the
clusters. However, the properties of the orbits are strongly affected
by the unknown distances of the Arches and the Quintuplet
clusters. Hence, strong conclusions are not possible until after it
becomes feasible to better constrain the distances.\looseness=-4

By combining the already-known clusters' positions and the
newly-computed absolute PMs, one important implication is clear
without the need for orbit calculations: the clusters not only both
lie close to the Disk plane, but also both move nearly parallel to
it. This strongly suggests orbits compatible with the general motion
of cold Galactic gas, as might be expected for young clusters, and
argues against more exotic origins of the clusters such as formation
in minor mergers or cooling halo gas. Assuming the same sense of
rotation as that of the gas, the motion towards positive $l$ (in a
frame where the GC is at rest) implies a location in front of the GC.

\section*{Acknowledgments}

ML, MF and AB acknowledge support from STScI grants GO 12915 and
13771. DJL acknowledges support from the Spanish Government Ministerio
de Ciencia, Innovaci\'on y Universidades through grants
PGC-2018-091\,3741-B-C22 and from the Canarian Agency for Research,
Innovation and Information Society (ACIISI), of the Canary Islands
Government, and the European Regional Development Fund (ERDF), under
grant with reference ProID2017010115. This work has made use of data
from the European Space Agency (ESA) mission {\it Gaia}
(\url{https://www.cosmos.esa.int/gaia}), processed by the {\it Gaia}
Data Processing and Analysis Consortium (DPAC,
\url{https://www.cosmos.esa.int/web/gaia/dpac/consortium}). Funding
for the DPAC has been provided by national institutions, in particular
the institutions participating in the {\it Gaia} Multilateral
Agreement.  This research made use of
Astropy,\footnote{\href{http://www.astropy.org}{http://www.astropy.org}}
a community-developed core Python package for Astronomy
\citep{2013A&A...558A..33A,2018AJ....156..123A}. This research has
made use of the SIMBAD database, operated at CDS, Strasbourg, France.

\appendix

\section{PM systematics}\label{appsys}

As described in Sect.~\ref{data}, bona-fide cluster members were used
by \citetalias{2015A&A...578A...4S} to compute the transformations
between frames in different epochs. The average PM of cluster stars
should be consistent with zero regardless of the magnitude and of the
position in the field of view of these objects.

We initially looked for possible magnitude-dependent systematics in
the PMs of \citetalias{2015A&A...578A...4S}. We computed the
3.5$\sigma$-clipped median value of the PM of cluster objects along
$\alpha \cos\delta$ and $\delta$ directions in 0.5-$K_{\rm S}$
magnitude bins. Cluster stars were defined by means of the membership
flag given in the \citetalias{2015A&A...578A...4S} catalogs. We also
refined the samples of members by considering as cluster stars only
those objects within 1 \masyr from the origin of the VPD.

We found a systematic PM offset for cluster stars with
$K_{\rm S} < 14$ with respect to the origin of the VPD that can be as
large as 0.5 \masyr in the central field of the Arches cluster. This
offset might be due to non-linearity effects in the NACO data, an
imperfect combination of long and short exposures \citepalias[which
happens at $K_{\rm S} \sim 14$ according to Fig.~3
of][]{2015A&A...578A...4S}, or simply because of a large contamination
of field stars among the bona-fide cluster members. We corrected the
PMs of \citetalias{2015A&A...578A...4S} by linearly interpolating
between the median-PM bin values.\looseness=-4

We also searched for spatially-variable systematics, i.e., local
offsets of the bulk motion of the cluster stars across the field of
view, and corrected for these effects as described in
\citet{2014ApJ...797..115B} and
\citet{2018ApJ...861...99L,2019ApJ...873..109L}. Most of the
systematics are located toward the center of the clusters and in
overlapping regions between the different fields.\looseness=-4

\begin{figure*}
  \centering
  \includegraphics[width=0.9\textwidth]{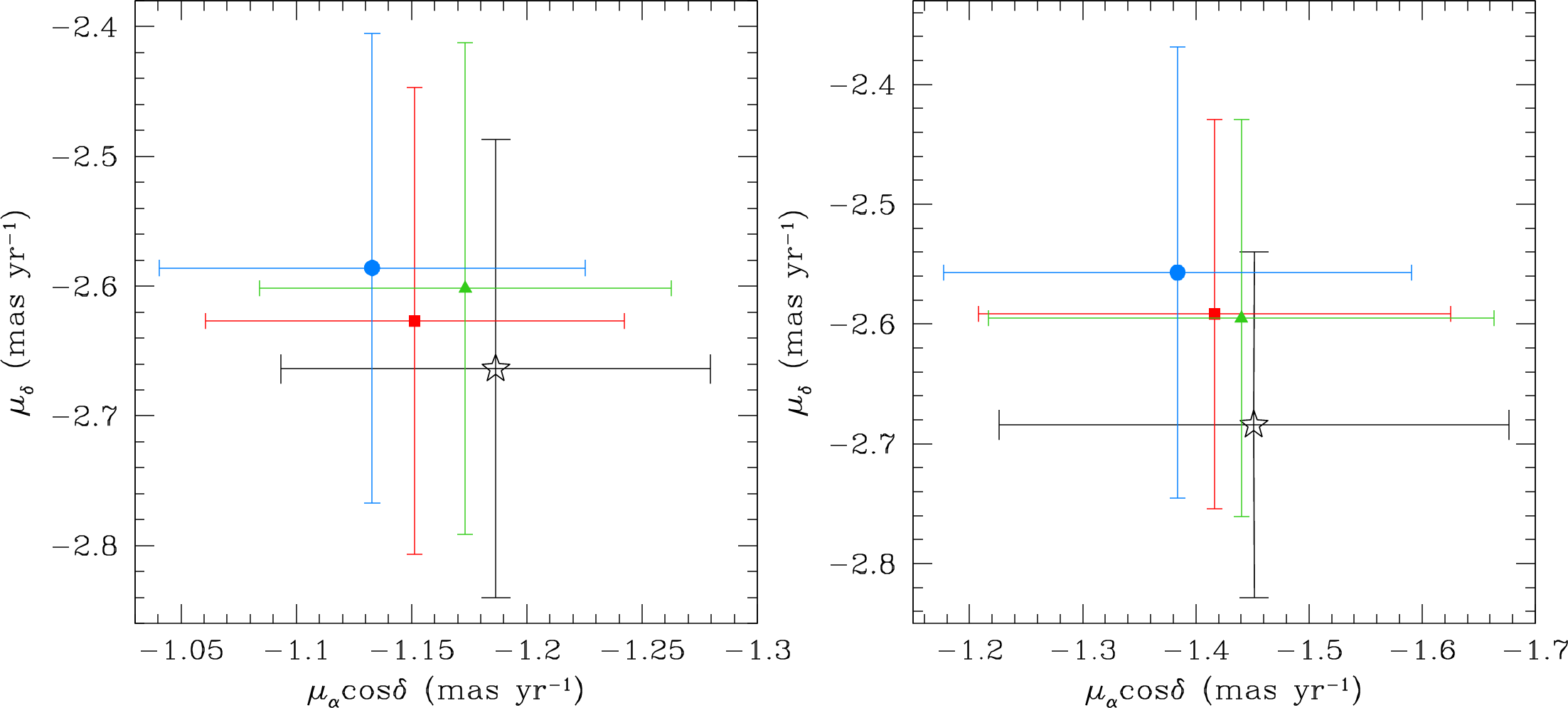}
  \caption{Comparison between the absolute PMs of the Quintuplet (left
    panel) and Arches (right panel) clusters obtained using the
    original PMs of \citetalias{2015A&A...578A...4S} (black stars),
    the PMs corrected for magnitude- (red squares) or
    spatial-dependent (green triangles) systematics, and the PMs
    corrected for both magnitude- and spatial-dependent systematic
    effects (azure dots).}
  \label{fig:corr}
\end{figure*}

In Fig.~\ref{fig:corr}, we show the absolute PM values of the
Quintuplet (left panel) and Arches (right panel) clusters obtained as
described in Sect.~\ref{gaia} by using: (i) the original PMs of
\citetalias{2015A&A...578A...4S} (black stars), (ii) the PMs corrected
for the magnitude-dependent systematics (red squares), (iii) the PMs
corrected for the spatially-variable systematics (green triangles),
and (iv) the PMs corrected for both magnitude- and spatial-dependent
systematic effects (azure dots).

It is clear that all measurements are in agreement within
$1\sigma$. Most of the stars in common with the Gaia DR2 catalog are
located outside the centermost fields, i.e., where the PM systematics
are larger. Therefore, the aforementioned corrections do not
significantly change the values of the absolute PMs of the two
clusters.

\section{The orbits of Arches and Quintuplet clusters}\label{orbit}

We briefly investigate the implications of our PM measurements by
calculating some possible orbits for the Arches and Quintuplet
clusters. Thanks to the addition of our PM results, five of the six
kinematic coordinates of each cluster are determined with reasonable
accuracy. The RVs adopted in this work are ($95 \pm 8$) km s$^{-1}$
for the Arches cluster \citep{2002ApJ...581..258F} and ($102 \pm 2$)
km s $^{-1}$ for the Quintuplet cluster
\citep{2014ApJ...789..115S}. The distance, however, is not well
determined. The heavy extinction toward the clusters makes any
distance estimates based on the brightness of cluster stars highly
uncertain. This distance uncertainty completely dominates the
uncertainty in our understanding of the orbits. The detailed
properties of the Galatic potential affect the orbits only to a lesser
degree. Therefore, we use here relatively simple axisymmetric models,
ignoring the influence of the Galactic bar, and leave more
sophisticated orbit modeling to a later time when the clusters'
distances are better known.\looseness=-4

To allow orbit computations over a wide range of radii, the model for
the Galactic potential we use here splices together two previously
published models.  In the inner region within 200 pc, we use a
spherically symmetric version of the potential of
\citet{2002A&A...384..112L}, which includes contributions from the
central black hole, the nuclear star cluster, the nuclear stellar
disk, and the Galactic bulge.  We use the mass profile of this model
as tabulated by \citet{2015MNRAS.447.1059K} and provided in the AMUSE
code \citep{2009NewA...14..369P}.  In the outer region, past a radius
of 200 pc, we use a modified version of the {\tt MWPotentialModel2014}
potential in \citet{2015ApJS..216...29B}.  While we keep the
Miyamoto-Nagai disk and Navarro-Frenk-White halo of this model
unchanged, we adopt a denser and more massive bulge in order to match
the level and slope of the \citeauthor{2002A&A...384..112L} rotation
curve at 200 pc.  We retain the functional form of the Bovy model's
bulge, $\rho(r) = \rho_s (r/r_1)^{-\alpha} \exp[-(r/r_c)^2]$, but with
revised parameters $\alpha = 2.0$, $r_c = 1.0$ kpc,
$\rho_s = 6.58 \times 10^{8} M_\odot$ kpc$^{-3}$, and $r_1 = 1$ kpc.
This yields a total bulge mass of $7.33 \times 10^9 M_\odot$.  The
model roughly matches the observed circular velocity data collected in
\citet{2013PASJ...65..118S}, for both large and small radii.  It does
not match this data's prominent peak at $R \sim 500$ pc and subsequent
decline out to 2.5 kpc, but \citet{2015A&A...578A..14C} argue that
this feature is likely an artifact of perturbations from the Galactic
bar.

The axisymmetry of our potential is clearly an approximation, as
orbits on scales of a few kpc are strongly affected by the bar.
However, at smaller radii, the observed physical components are close
to axisymmetric and the potential contours generated in physical
models such as \citet{2003MNRAS.340..949B} are more rounded.  Thus,
axisymmetric models continue to be used to study orbits near the
Galactic center
\citep{2011ApJ...735L..33M,2015MNRAS.447.1059K,2019MNRAS.487.1025P},
and we leave non-axisymmetric refinements to future work.

We integrate orbits in this potential using the package {\tt
  galpy}\footnote{\href{http://github.com/jobovy/galpy}{http://github.com/jobovy/galpy}}
\citep{2015ApJS..216...29B}. We assume Sgr A* lies fixed at the GC.
We adopt a distance to Sgr A* of 8.175 kpc from
\citet{2019A&A...625L..10G}, and do not consider in the computation
its small observed error. We first define a modified Galactic
coordinate system $(l', b')$, which uses a slight rotation around the
Sun's position to place Sgr A* exactly at $l' = 0$, $b' = 0$, then
shift by the solar distance to obtain Galactocentric coordinates. We
choose a solar motion in the Galactocentric frame of $v_X = -11.0$ km
s$^{-1}$, $v_Y = 248.5$ km s$^{-1}$, $v_Z = 7.8$ km s$^{-1}$ that
produces the observed PM of Sgr A*, $\mu_\alpha \cos\delta = -3.156$
\masyr and $\mu_\delta = -5.585$ \masyr
\citep{2020ApJ...892...39R}. We use a Galactocentric convention where
$X$ points from Sgr A* to the Sun.

\begin{figure*}
  \includegraphics[width=0.9\textwidth]{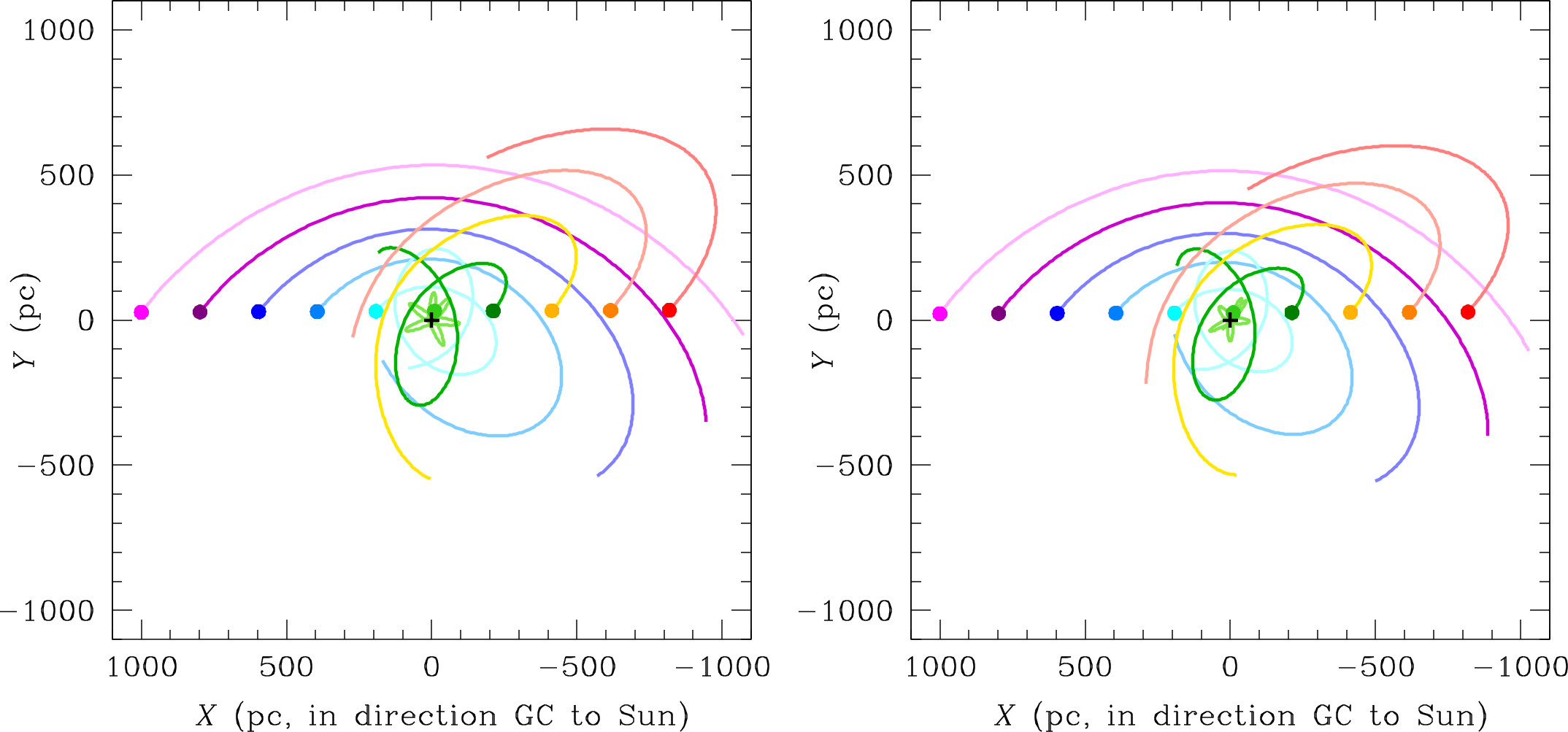}
  \caption{Orbits in the axisymmetric potential for the Quintuplet
    (left) and Arches (right) clusters. For each cluster, the
    centermost starting point is set at the same distance as that of
    the GC. Five orbits are also simulated for distances
    closer/further than the GC with steps of 0.2 kpc.  For clarity, we
    show here only the orbits integrated forward in time for 10
    Myr. The position of Sgr A* is marked with a black cross.}
  \label{fig.axi_multi}
\end{figure*}

Figure~\ref{fig.axi_multi} shows orbits of the two clusters with a
series of initial distances, using our best values of the PM.  Orbits
starting well on the near side of the GC are highly eccentric. The
eccentricity first decreases with the starting distance, then
increases again as the orbits become almost radial near the GC.  About
20 pc further out, the orbits become retrograde.  We note none of the
orbits are circular.  As noted in \citet{2008ApJ...675.1278S},
circular orbits require $\mathbf{r} \cdot \mathbf{v} = 0$, which
implies $X = (v_X / v_Y) Y$, radius $R = [1 + (v_X / v_Y)^2]^{1/2} Y$,
and circular velocity $V_c = (v_X^2 + v_Y^2)^{1/2}$.  For Quintuplet,
this implies a velocity $V_c = 177 \pm 4$ km s$^{-1}$ at $R=48 \pm 1$
pc, where the uncertainties incorporate our estimated PM and RV
errors. For Arches, it implies a velocity $V_c = 168 \pm 7$ km
s$^{-1}$ at $R=40 \pm 2$ pc. Both of these circular velocities are
comfortably above those implied by our potential, which are 108 and 98
km s$^{-1}$, respectively, at these radii.  They also lie well above
the scatter of different observational points
\citep{2002A&A...384..112L,2013PASJ...65..118S}.  Therefore, even
though our PM values imply transverse speeds that are somewhat smaller
for Quintuplet than those in \citet{2014ApJ...789..115S}, and quite a
bit smaller for Arches than in \citet{2008ApJ...675.1278S}, we agree
with the assessment in those papers that neither cluster lies on a
circular orbit.

\begin{figure*}
  \includegraphics[width=0.9\textwidth]{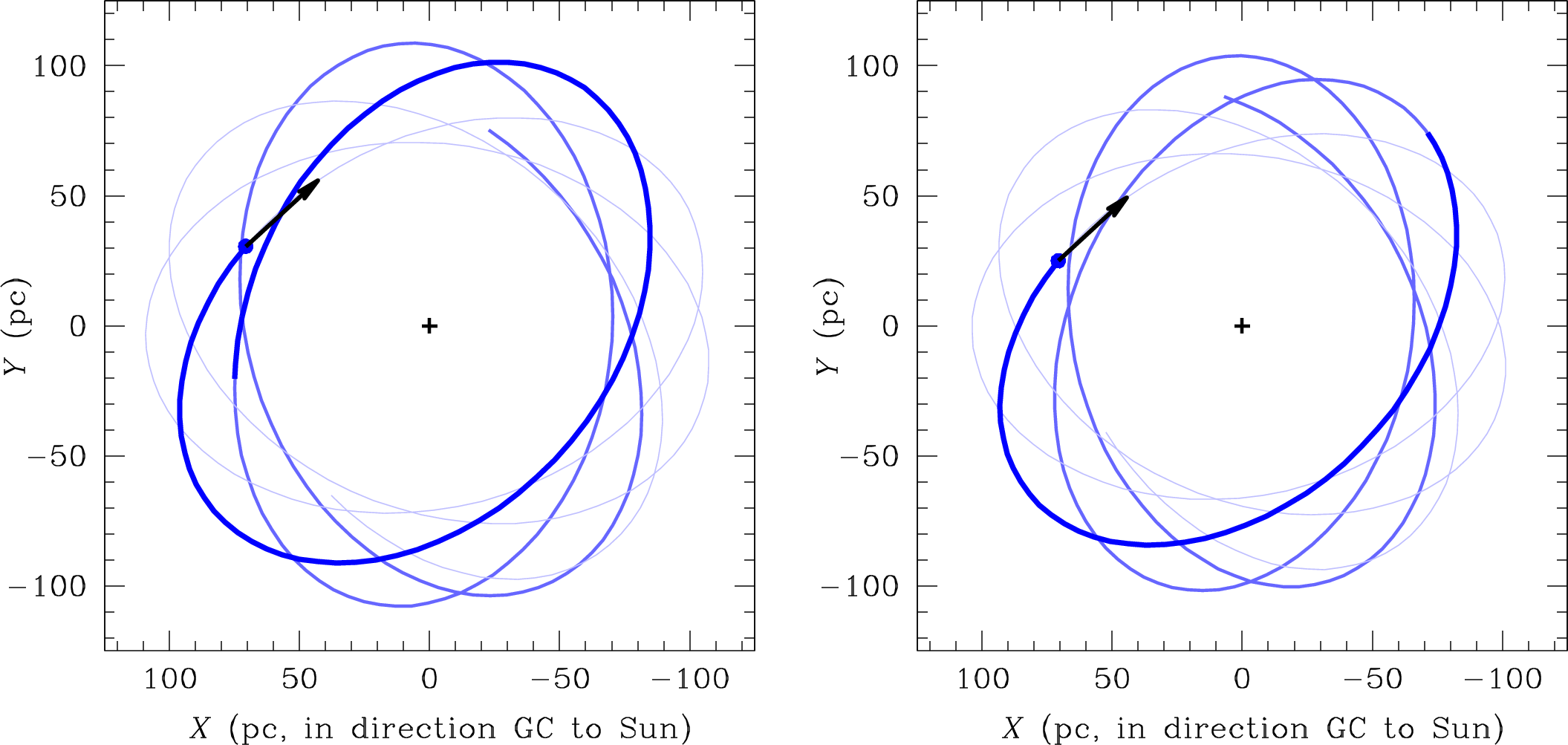}
  \caption{Orbits of minimum eccentricity in the axisymmetric
    potential for the Quintuplet (left) and Arches (right)
    clusters. The heaviest line traces backward for the assumed age of
    each cluster, while further continuations forward and backward are
    provided to illustrate the character of the orbit. The position of
    Sgr A* in both panels is marked with a black cross.}
  \label{fig.axi}
\end{figure*}

Our orbits are rather similar in character to those in \citet[see
their Figure 8]{2014ApJ...789..115S}.  One possible reason for this is
that their bar component \citep[as specified
in][]{2014ApJ...789..115S} only becomes strong past 200 pc, leaving
the central portion nearly axisymmetric.  If we presume the clusters
to have been formed out of gas in circular rotation, the orbits with
the minimum eccentricity may be regarded as the most plausible.  These
are shown in the two panels in Figure~\ref{fig.axi}.  For Arches, this
orbit has apocenter 105 pc, pericenter 66 pc, eccentricity
$\equiv (R_{\rm max}-R_{\rm min})/(R_{\rm max}+R_{\rm min}) = 0.23$,
and radial period 2.0 Myr.  For Quintuplet, these numbers change to
apocenter 110 pc, pericenter 71 pc, eccentricity 0.22, with radial
period 2.0 Myr.  The two orbits are thus remarkably similar.  These
numbers are not drastically changed by variations up to $2 \sigma$ in
the PM.

The orbit properties are affected above all by the uncertainty in the
distance, but we can make some arguments to constrain the plausible
range of this parameter. One such argument asserts that the current
values of Galactocentric coordinates should not be special. The
projected locations of the clusters are 27 and 31 pc away from the
GC\footnote{Positions from the
  \href{http://simbad.u-strasbg.fr/simbad/}{Simbad} database
  \citep{2000A&AS..143....9W}.} in the $Y$ direction for the Arches
and Quintuplet clusters, respectively.  We exclude here any orbit for
which $R$ or $|Y|$ are smaller than the current values less than 5\%
of the time.  We also exclude retrograde orbits, since these young
clusters are likely to have been born within prograde gas.

Orbits with a slightly greater starting distance become quite radial
as the velocity vector begins to align with the Galactocentric
position vector.  These take the clusters onto highly disruptive
orbits. Given cluster masses of $\sim 10^4 M_\odot$ and half-mass
radii of $\sim 0.4$ pc
\citep{2002ApJ...565..265P,2008ApJ...675.1278S}, the enclosed Galactic
tidal force exceeds the cluster self-gravity within $\sim$15--20 pc,
which should lead to rapid tidal disruption.  It also seems unlikely
that star-forming gas suitable for forming these young clusters would
be located on this type of orbit.  This argues against cluster
apocenter-pericenter ratios of greater than about 5:1 for our sample
of orbits.

For both Arches and Quintuplet, application of these criteria limits
the range of the current $X$ coordinate to roughly $10$--$350$ pc,
where the cluster lies on the near side of the GC.  This yields orbits
with apocenters of $80$--$450$ pc and radial periods of roughly
$1.7$--$7$ Myr.

For some orbits, the clusters complete less than one full radial
period.  \citet{2002ApJ...565..265P} argued that clusters near the GC
rapidly decrease in surface density due to tidal losses, which may
help explain why such young clusters are the only ones observed near
the GC, and also favor orbits with fewer and larger pericenters.  The
close similarity in position and velocity suggests the clusters may
actually originate from the same material. While the difference in
their estimated ages might suggest otherwise, these ages are highly
uncertain. Furthermore, molecular clouds are variously estimated to
live for as long as $\sim 30$ Myr \citep{2011ApJ...729..133M}, though
close to the GC they may have shorter lifetimes of 3--9 Myr
\citep{2018MNRAS.478.3380J} and clouds may disperse within 1--5 Myr
once massive stars form \citep{2020MNRAS.493.2872C}. Therefore, it
might not be impossible that the clusters could have formed out of
different parts of the same large gas cloud.  Indeed, it is not
difficult to find orbits where they originate from the same point 4
Myr ago, our assumed age for Quintuplet.  However, given their
estimated masses, it is clear that the clusters are not {\it
  currently} gravitationally bound, as their projected separation is
greater than the Jacobi radius implied by their combined mass and any
pericentric radius up to $\sim 500$ pc. If the clusters are on orbits
that drift away from each other only slowly with time, however, this
not only could explain their apparently similar azimuth, but also
makes it much less of a coincidence they are both projected so close
to the GC.

\section*{Data availability}

The catalogs of \citet{2015A&A...578A...4S} are available at
\href{http://cdsarc.u-strasbg.fr/viz-bin/qcat?J/A+A/578/A4}{http://cdsarc.u-strasbg.fr/viz-bin/qcat?J/A+A/578/A4}. The
Gaia DR2 catalog is available at
\href{https://gea.esac.esa.int/archive/}{https://gea.esac.esa.int/archive/}.

\bibliographystyle{mnras}

\begin{thebibliography}{}

\bibitem[\protect\citeauthoryear{Astropy Collaboration et al.}{2013}]{2013A&A...558A..33A} Astropy Collaboration et al., 2013, A\&A, 558, A33

\bibitem[\protect\citeauthoryear{Astropy Collaboration et al.}{2018}]{2018AJ....156..123A} Astropy Collaboration et al., 2018, AJ, 156, 123

\bibitem[\protect\citeauthoryear{Bellini et al.}{2014}]{2014ApJ...797..115B} Bellini A. et al., 2014, ApJ, 797, 115

\bibitem[\protect\citeauthoryear{Bissantz, Englmaier \& Gerhard}{2003}]{2003MNRAS.340..949B} Bissantz N., Englmaier P., Gerhard O., 2003, MNRAS, 340, 949

\bibitem[\protect\citeauthoryear{Bovy}{2015}]{2015ApJS..216...29B} Bovy J., 2015, ApJS, 216, 29

\bibitem[\protect\citeauthoryear{Chemin et al.}{2015}]{2015A&A...578A..14C} Chemin, L., Renaud, F., \& Soubiran, C.\ 2015, \aap, 578, A14

\bibitem[\protect\citeauthoryear{Chevance et al.}{2020}]{2020MNRAS.493.2872C} Chevance M., Kruijssen J.~M.~D., Hygate A.~P.~S., Schruba A., Longmore S.~N., Groves B., Henshaw J.~D., et al., 2020, MNRAS, 493, 2872

\bibitem[\protect\citeauthoryear{Clarkson et al.}{2012}]{2012ApJ...751..132C} Clarkson W.~I. et al., 2012, ApJ, 751, 132

\bibitem[\protect\citeauthoryear{Figer, McLean \& Morris}{1999a}]{1999ApJ...514..202F} Figer D.~F., McLean I.~S., Morris M., 1999a, ApJ, 514, 202

\bibitem[\protect\citeauthoryear{Figer et al.}{1999b}]{1999ApJ...525..750F} Figer D.~F., Kim S.~S., Morris M., Serabyn E., Rich R.~M., McLean I.~S., 1999b, ApJ, 525, 750

\bibitem[\protect\citeauthoryear{Figer et al.}{2002}]{2002ApJ...581..258F} Figer D.~F. et al., 2002, ApJ, 581, 258

\bibitem[\protect\citeauthoryear{Fruchter \& Hook}{2002}]{2002PASP..114..144F} Fruchter A.~S., Hook R.~N., 2002, PASP, 114, 144

\bibitem[\protect\citeauthoryear{Gaia Collaboration et al.}{2016}]{2016A&A...595A...1G} Gaia Collaboration et al., 2016, A\&A, 595, A1
  
\bibitem[\protect\citeauthoryear{Gaia Collaboration et al.}{2018}]{2018A&A...616A...1G} Gaia Collaboration et al., 2018, A\&A, 616, A1
  
\bibitem[\protect\citeauthoryear{Genzel, Eisenhauer \& Gillessen}{2010}]{2010RvMP...82.3121G} Genzel R., Eisenhauer F., Gillessen S., 2010, RvMP, 82, 3121

\bibitem[\protect\citeauthoryear{Gravity Collaboration et al.}{2019}]{2019A&A...625L..10G} Gravity Collaboration et al., 2019, A\&A, 625, L10

\bibitem[\protect\citeauthoryear{Harfst, Portegies Zwart \& Stolte}{2010}]{2010MNRAS.409..628H} Harfst S., Portegies Zwart S., Stolte A., 2010, MNRAS, 409, 628

\bibitem[\protect\citeauthoryear{Habibi}{2014}]{2014PhDT.......214H} Habibi M., 2014, PhDT

\bibitem[\protect\citeauthoryear{Hosek et al.}{2019}]{2019ApJ...870...44H} Hosek M.~W. et al., 2019, ApJ, 870, 44

\bibitem[\protect\citeauthoryear{Jeffreson et al.}{2018}]{2018MNRAS.478.3380J} Jeffreson S.~M.~R., Kruijssen J.~M.~D., Krumholz M.~R., Longmore S.~N., 2018, MNRAS, 478, 3380

\bibitem[\protect\citeauthoryear{Kruijssen et al.}{2015}]{2015MNRAS.447.1059K} Kruijssen, J.~M.~D., Dale, J.~E., \& Longmore, S.~N.\ 2015, \mnras, 447, 1059

\bibitem[\protect\citeauthoryear{Launhardt, Zylka \& Mezger}{2002}]{2002A&A...384..112L} Launhardt R., Zylka R., Mezger P.~G., 2002, A\&A, 384, 112

\bibitem[\protect\citeauthoryear{Libralato et al.}{2018a}]{2018ApJ...854...45L} Libralato M. et al., 2018a, ApJ, 854, 45

\bibitem[\protect\citeauthoryear{Libralato et al.}{2018b}]{2018ApJ...861...99L} Libralato M. et al., 2018b, ApJ, 861, 99

\bibitem[\protect\citeauthoryear{Libralato et al.}{2019}]{2019ApJ...873..109L} Libralato M. et al., 2019, ApJ, 873, 109

\bibitem[\protect\citeauthoryear{Lindegren et al.}{2018}]{2018A&A...616A...2L} Lindegren L. et al., 2018, A\&A, 616, A2

\bibitem[\protect\citeauthoryear{Lu}{2018}]{2018ASSL..424...69L} Lu J.~R., 2018, ASSL,  69, ASSL..424

\bibitem[\protect\citeauthoryear{Molinari et al.}{2011}]{2011ApJ...735L..33M} Molinari S. et al., 2011, ApJL, 735, L33

\bibitem[\protect\citeauthoryear{Murray}{2011}]{2011ApJ...729..133M} Murray N., 2011, ApJ, 729, 133

\bibitem[\protect\citeauthoryear{Najarro et al.}{2004}]{2004ApJ...611L.105N} Najarro F., Figer D.~F., Hillier D.~J., Kudritzki R.~P., 2004, ApJL, 611, L105

\bibitem[\protect\citeauthoryear{Plewa et al.}{2015}]{2015MNRAS.453.3234P} Plewa P.~M. et al., 2015, MNRAS, 453, 3234

\bibitem[\protect\citeauthoryear{Perera et al.}{2019}]{2019MNRAS.487.1025P} Perera B.~B.~P. et al., 2019, MNRAS, 487, 1025

\bibitem[\protect\citeauthoryear{Portegies Zwart et al.}{2002}]{2002ApJ...565..265P} Portegies Zwart S.~F., Makino J., McMillan S.~L.~W., Hut P., 2002, ApJ, 565, 265

 \bibitem[\protect\citeauthoryear{Portegies Zwart et al.}{2009}]{2009NewA...14..369P} Portegies Zwart S. et al., 2009, NewA, 14, 369

\bibitem[\protect\citeauthoryear{Reid \& Brunthaler}{2020}]{2020ApJ...892...39R} Reid M.~J., Brunthaler A., 2020, ApJ, 892, 39

\bibitem[\protect\citeauthoryear{Rui et al.}{2019}]{2019ApJ...877...37R} Rui N.~Z., Hosek M.~W., Lu J.~R., Clarkson W.~I., Anderson J., Morris M.~R., Ghez A.~M., 2019, ApJ, 877, 37

\bibitem[\protect\citeauthoryear{Sofue}{2013}]{2013PASJ...65..118S} Sofue Y., 2013, PASJ, 65, 118

\bibitem[\protect\citeauthoryear{Stolte et al.}{2008}]{2008ApJ...675.1278S} Stolte A., Ghez A.~M., Morris M., Lu J.~R., Brandner W., Matthews K., 2008, ApJ, 675, 1278

\bibitem[\protect\citeauthoryear{Stolte et al.}{2014}]{2014ApJ...789..115S} Stolte A. et al., 2014, ApJ, 789, 115

\bibitem[\protect\citeauthoryear{Stolte et al.}{2015}]{2015A&A...578A...4S} Stolte A. et al., 2015, A\&A, 578, A4, S15

\bibitem[\protect\citeauthoryear{Wenger et al.}{2000}]{2000A&AS..143....9W} Wenger M. et al., 2000, A\&AS, 143, 9

\end{thebibliography}

\bsp	
\label{lastpage}
\end{document}